# MgAl$_2$O$_4$:Cr$^{3+}$ and emerald display a different colour but the local symmetry is the same: Microscopic origin

J. M. García-Lastra[1], M. T. Barriuso[2], J. A. Aramburu[3], and M. Moreno[3]

[1]*Departamento de Física de Materiales, Facultad de Químicas, Universidad del País Vasco, 20018 San Sebastián, Spain*

[2]*Departamento de Física Moderna, Universidad de Cantabria, 39005 Santander, Spain*

[3]*Departamento de Ciencias de la Tierra y Física de la Materia Condensada, Universidad de Cantabria, 39005 Santander, Spain*

**Abstract**

The difference in colour between emerald (Be$_3$Si$_6$Al$_2$O$_{18}$:Cr$^{3+}$, green) and the Cr$^{3+}$-doped spinel MgAl$_2$O$_4$ (red) is striking, considering that in both systems colour is due to CrO$_6^{9-}$ complexes with D$_3$ symmetry and the measured Cr$^{3+}$-O$^{2-}$ distance is practically the same (1.98 ± 0.01 and 1.97 ± 0.01 Å, respectively). By means of density functional calculations it is shown that this surprising difference can reasonably be explained once the electric field, **E$_R$**, which all lattice ions lying *outside* the CrO$_6^{9-}$ complex exert on localized electrons, is taken into consideration. The origin of the different shape of **E$_R$** in the two host lattices is analysed in detail. It is shown that **E$_R$** raises (decreases) the 2p(O) levels for Be$_3$Si$_6$Al$_2$O$_{18}$:Cr$^{3+}$ (MgAl$_2$O$_4$:Cr$^{3+}$) along the trigonal axis thus favouring a decrease (increase) of 10Dq. The present work demonstrates the key role played by **E$_R$** (not considered in the traditional ligand field theory) for understanding the differences exhibited by the same complex embedded in host lattices which do not have the same crystal structure. Some remarks on the colour of Cr$_2$O$_3$ pure compound are also reported.

# 1. Introduction

A great deal of experimental work has been devoted to look into the properties of gemstones and minerals involving $Cr^{3+}$-doped oxides[1-3]. Despite these efforts the actual origin of the colour exhibited by the different oxides lattices containing $Cr^{3+}$ is still being debated[3-10].

Historically, the optical properties of gemstones like ruby ($Cr^{3+}$-doped corundum, $Al_2O_3$) and emerald ($Cr^{3+}$-doped beryl, $Be_3Si_6Al_2O_{18}$) have been interpreted in the framework of the traditional ligand field theory (LFT)[11,12]. In this domain, it is assumed that the electronic properties of a transition metal impurity, M, placed in an insulating lattice can be understood considering *only* the $MX_N$ complex formed with the N nearest ions or ligands, X. The fact that crystal-field spectra of $KMF_3$ (M = Ni, Mn) pure compounds look very similar to those measured for $KMgF_3:M^{2+}$ (M = Ni, Mn) support such an assumption[13-15]. Along this line theoretical calculations on $NiF_6^{4-}$, $MnF_6^{4-}$, $CrF_6^{3-}$ or $CrO_4^{4-}$ complexes *in vacuo* at the right experimental distance[16-20] is known to give values of optical transitions and electron paramagnetic resonance (EPR) parameters which are not far from experimental data[13-15,21,22] for $KMgF_3:M^{2+}$ (M = Ni, Mn), $K_2NaGaF_6:Cr^{3+}$ or $Mg_2SiO_4: Cr^{4+}$.

If the idea of complex is fully right the optical properties of an octahedral $MX_N$ complex embedded in a series of different host lattices would depend only on the actual value of the *equilibrium* metal-ligand distance. This statement has been verified to be right looking at the different optical spectra of $MnF_6^{4-}$, $NiF_6^{4-}$ and $CrF_6^{3-}$ complexes located in a series of distinct but *isomorphous* host lattices[23-25].

By virtue of these facts the red and green colour exhibited by ruby and emerald, respectively, have often been ascribed to a different value of the mean equilibrium $Cr^{3+}$-$O^{2-}$ distance, $R_I$, in the $CrO_6^{9-}$ complex[1,2,26]. Recent extended X-ray absorption fine structure (EXAFS) measurements carried out on ruby and emerald prove[3,4,6,9] however that the $R_I$ value for both gemstones is the same within $\pm 0.01$ Å.

The colour of insulating oxides doped with $Cr^{3+}$ essentially depends on the energy of the first spin allowed transition $^4A_{2g}$ $(t_{2g}^3)$ $\rightarrow$ $^4T_{2g}$ $(t_{2g}^2e_g)$ transition[11] which is just equal to the cubic field splitting parameter, 10Dq. Within the traditional LFT it is assumed that

$$10Dq = (10Dq)_v \tag{1}$$

where $(10Dq)_v$ stands for the complex in vacuo. Furthermore, it is assumed that $(10Dq)_v$ for a given complex only depends on the metal-ligand distance, R, through the law

$$10Dq = CR^{-n} \tag{2}$$

where C is a constant. Optical absorption measurements under hydrostatic pressure carried out for $Al_2O_3:Cr^{3+}$ have shown[27] that experimental 10Dq values are reproduced by Eqs. (1) and (2) with $n = 4.5$. Similar results have been obtained for other transition metal complexes[12,15]. The microscopic origin of the strong dependence of $(10Dq)_v$ upon R has previously been discussed[12,19,28,29]. Bearing in mind these facts it is thus not possible to conciliate the optical and structural data for ruby and emerald with the usual view provided by the LFT.

A solution for understanding this somewhat puzzling situation has recently been put forward[5,7,10]. It has been argued that even if active electrons are well localized in the complex it should also been taken into consideration the electric field, $\mathbf{E}_R$, created by all ions of the insulating lattice lying *outside* the complex, on the electrons *in* the complex. The internal field $\mathbf{E}_R$ should, in principle, be present for every insulating lattice composed by ions. This means that all properties (and thus 10Dq) associated with a complex do also depend on the shape of $\mathbf{E}_R$ in the complex region. In particular it has been emphasized the importance of incorporating $\mathbf{E}_R(\mathbf{r})$ when comparing the properties of the same complex placed in two host lattices which are *not isomorphous*.

According to this new standpoint a difference between ruby and emerald comes out merely considering the local symmetry around the $Cr^{3+}$ impurity. In fact, in $Al_2O_3:Cr^{3+}$ the local symmetry is $C_3$ and thus there is an electric field at the chromium site placed at $\mathbf{r} = 0$. However, in the case of emerald $\mathbf{E}_R(0)$ is rigorously null as a result of a higher local symmetry ($D_3$).

A more subtle problem has recently been raised[8] in the comparison of emerald ($Be_3Si_6Al_2O_{18}:Cr^{3+}$) with the spinel $MgAl_2O_4$ doped with $Cr^{3+}$. In both cases the $Cr^{3+}$ impurity enters the $Al^{3+}$ site and the symmetry of the $CrO_6^{9-}$ complex is $D_3$. Recent EXAFS measurements[8,9] have lead to a $R_I = 1.97 \pm 0.01$ Å value for emerald while $R_I = 1.98 \pm 0.01$ Å for $MgAl_2O_4:Cr^{3+}$. In spite of these facts the energy of the first spin allowed $^4A_{2g}$ ($t_{2g}^3$) $\rightarrow$ $^4T_{2g}$ ($t_{2g}^2e_g$) transition has been measured[1,30,31] to be equal to 18520 cm$^{-1}$ for $MgAl_2O_4:Cr^{3+}$ while it is equal only to 16130 cm$^{-1}$ for $Be_3Si_6Al_2O_{18}:Cr^{3+}$. This means that, although emerald and the spinel $MgAl_2O_4$ doped with $Cr^{3+}$ share the same local symmetry and have practically the same $R_I$ value, the colour displayed by $MgAl_2O_4:Cr^{3+}$ is red (identical to that of ruby for the human eye[26]) and not green.

The present work is aimed at clarifying this relevant issue by means of the same procedure previously employed[5,7] in the study of ruby, emerald and the two centres (with $C_s$ and $C_i$ symmetries) formed in alexandrite ($BeAl_2O_4:Cr^{3+}$). Accordingly, 10Dq is derived by means of density functional calculations, considering the $CrO_6^{9-}$ complex at the right equilibrium geometry and subject to the internal field, $\mathbf{E_R}(\mathbf{r})$, coming from the $MgAl_2O_4$ host lattice. For well clearing out the origin of differences between $MgAl_2O_4:Cr^{3+}$ and emerald particular attention is paid to look into the shape of $\mathbf{E_R}(\mathbf{r})$ in the two $MgAl_2O_4$ and $Be_3Si_6Al_2O_{18}$ host lattices.

## 2. Computational Details

Calculations have been performed in the framework of the density functional theory (DFT) by means of the Amsterdam density functional (ADF) code[32]. All results shown in this paper have been performed on $CrO_6^{9-}$ clusters at their experimental equilibrium geometries. For both systems, 10Dq has been computed for the complex *in vacuo* as well as including the effects of the electrostatic potential, $V_R(\mathbf{r})$, generating $\mathbf{E_R}$ through the relation $\mathbf{E_R}(\mathbf{r}) = -\nabla V_R(\mathbf{r})$. The effects of $V_R(\mathbf{r})$ have been included by means of the same technique described in previous works[5,7].

The same functional and basis set are employed for calculating the emerald and the spinel. The generalized gradient approximation (GGA) exchange-correlation energy was computed using the Perdew-Wang functional[33], PW91. It was verified that main results obtained in the present calculations are almost independent on the used functional. The $Cr^{3+}$ ion has been described through basis sets of TZP (triple-$\zeta$ Slater-type orbitals STO plus one polarization function) quality given in the program database, keeping the core electrons (1$s$-3$p$) frozen. In the case of $O^{2-}$ ions, a DZP (double-$\zeta$ Slater-type orbitals STO plus one polarization function) basis sets quality has been used, keeping the 1s shell frozen. This is the description for oxygen ions which has provided better agreement with experimental findings in recent works[5,7,34].

The 10Dq parameter has been derived following the average of configuration procedure[35] based on Slater´s transition state concept[36]. In the case of cubic symmetry the Kohn-Sham equations are solved for the $t_{2g}^{9/5}e_g^{6/5}$ configuration where all mainly $d$-levels are equally populated. As shown in Ref. [35], the difference between the $\varepsilon(e_g)$ and $\varepsilon(t_{2g})$ eigenvalues derived for such configuration with fractional occupation leads to a reasonable 10Dq value. This procedure can easily be extended if the symmetry of the complex is lower than $O_h$, such as it happens in the present cases.

## 3. Results and Discussion

Seeking to look into the influence of the internal $\mathbf{E}_R(\mathbf{r})$ field on 10Dq and the colour of $MgAl_2O_4$:$Cr^{3+}$ and emerald gemstones, calculations have been carried out in two steps. In the first one, 10Dq has been derived for the $D_3$ $CrO_6^{9-}$ complex *in vacuo* at the experimental equilibrium geometry thus considering the effects of small trigonal distortions. In a second step, the action of the internal electric field, $\mathbf{E}_R(\mathbf{r})$, upon active electrons in the complex is incorporated into the calculation. Main results are collected in Table 1. In order to show the strong dependence of 10Dq upon R, calculated values for $MgAl_2O_4$:$Cr^{3+}$ at a distance R = 1.995 Å, very close to the experimental figure[8] $R_I$ = 1.98 ± 0.01 Å, have also been included in Table 1. For the sake of completeness in that table results for ruby are also reported, while values of the average $Al^{3+}$-$O^{2-}$ distance in the perfect host lattices[4,8,37], $R_H$, are enclosed for comparison purposes.

It is worth noting that for emerald and the spinel the six $Cr^{3+}$-$O^{2-}$ distances are equal although $O^{2-}$-$Cr^{3+}$-$O^{2-}$ angles do not correspond to a perfect octahedron[4,8,9,37]. The departure from octahedral geometry is bigger for emerald than for $MgAl_2O_4$:$Cr^{3+}$. For instance, for two $O^{2-}$ ions in trans position the $O^{2-}$-$Cr^{3+}$-$O^{2-}$ angle is equal to 170.5º in the case of emerald while equal to 180º for spinel. The existence of this trigonal distortion in the $CrO_6^{9-}$ complex leads to small differences in the calculated 10Dq values for emerald and $MgAl_2O_4$:$Cr^{3+}$ when only the complex in vacuo is considered. Let us call $\Delta_{SE} = 10Dq(MgAl_2O_4$:$Cr^{3+}) - 10Dq(emerald)$. As shown in Table 1 the calculated value for the complex in vacuo is $\Delta_{SE} \sim 300$ cm$^{-1}$ and thus it is eight times smaller than the experimental value $\Delta_{SE} = 2390$ cm$^{-1}$.

Looking at the results gathered in Table 1 it is also hard to understand the distinct 10Dq values exhibited by $MgAl_2O_4$:$Cr^{3+}$ and emerald through the complex in vacuo even if the uncertainty in the experimental $R_I$ value ($\pm$ 0.01 Å) is considered. In fact, as $R_H = 1.93$ Å for the spinel while $R_H = 1.906$ Å for beryl, it can reasonably be expected that $R_I(MgAl_2O_4$:$Cr^{3+}) \geq R_I$ (emerald) according to the general behaviour observed when a given complex is inserted in different host lattices[12,23]. In fact, for $3d$ complexes placed in a series of cubic insulating lattices it has been found that $R_I$ values are ordered in the same way as $R_H$. Bearing in mind these facts, Eq. (1) and the results of Table 1 it can be concluded that if we only consider the complex in vacuo $\Delta_{SE}$ is expected to be smaller than 300 cm$^{-1}$.

As shown in Table 1, a significant *increase* on the calculated 10Dq value of $MgAl_2O_4$:$Cr^{3+}$ is obtained once the corresponding internal electric field, $\mathbf{E}_R(\mathbf{r})$, is incorporated into the calculation. In agreement with what was previously reported[5,7], $\mathbf{E}_R(\mathbf{r})$ is found to *reduce* but only by ~450 cm$^{-1}$ the 10Dq value derived for emerald using a complex in vacuo. Therefore, the variation on 10Dq induced by $\mathbf{E}_R(\mathbf{r})$ in this gemstone has a different sign to that in $MgAl_2O_4$:$Cr^{3+}$. These results just mean that when the complex is inserted in a lattice there is a *supplementary* contribution to 10Dq coming from $V_R(\mathbf{r})$, termed as $\Delta_R$, and Eq. (1) has to be modified to[10]

$$10Dq = (10Dq)_v + \Delta_R \tag{3}$$

It has recently been pointed out that $\Delta_R$ plays a relevant role even if host lattices are cubic[38]. The results gathered in Table 1 then support that the internal $\mathbf{E}_R(\mathbf{r})$ field plays a key role for understanding why $MgAl_2O_4$:$Cr^{3+}$ is red despite the local symmetry around $Cr^{3+}$ is $D_3$, the same as for emerald. Despite this fact the results embodied in Table 1 and those previously obtained[5,7,38] stress that the main contribution to 10Dq comes from $(10Dq)_v$.

An insight into the origin of such a difference can be gained looking at Fig. 1, where the arrangement of neighbour ions to the $CrO_6^{9-}$ complex can be seen for both $MgAl_2O_4$ and $Be_3Si_6Al_2O_{18}$ host lattices. In both cases the direction called $d_1$ in Fig. 1 corresponds to the $C_3$ axis in $D_3$ symmetry. Despite the local symmetry around $Cr^{3+}$ is the same in both lattices the nature and arrangement of first and second shell of ions looks certainly different. In the case of $MgAl_2O_4$ the second shell around the central $Al_c^{3+}$ ion (which is replaced by the $Cr^{3+}$ impurity) is composed[37] by six $Al^{3+}$ ions placed at 2.86 Å, while the third shell is formed by two $O^{2-}$ ions lying at 3.33 Å. All these ions are shown in Fig. 1. Next there are six $Mg^{2+}$ located at 3.35 Å and six $O^{2-}$ ions at 3.56 Å while further shells are all lying at a distance higher than 4 Å. Differences between the local geometry in $MgAl_2O_4$ and $Be_3Si_6Al_2O_{18}$ are already visible looking at the second shell. In fact, in $Be_3Si_6Al_2O_{18}$ there are only three $Be^{3+}$ ions lying at 2.66 Å from the central $Al_c^{3+}$ ion involved in this shell[4,9]. Further differences between two lattices appear considering the angle, $\phi_3$, formed by an $Al_c^{3+}$ - $M_{2s}^{3+}$ direction with the principal $C_3$ axis. Here $M_{2s}^{3+}$ just means a cation of the second shell. While $\phi_3 = 90°$ for beryl lattice, $\phi_3$ is equal only to 35.26° for the spinel. As shown in Fig. 1 six $Si^{4+}$ ions, at 3.28 Å from $Al_c^{3+}$, are involved in the third shell of $Be_3Si_6Al_2O_{18}$. The fourth sell is composed by six $O^{2-}$ ions at 3.73 Å from $Al_c^{3+}$.

Bearing in mind the structural differences between the spinel and the beryl lattices let us now have a look to the form of the calculated $\mathbf{E}_R$ field in the two lattices. For seeing in what places of the complex region there is an electric field $\mathbf{E}_R(\mathbf{r}) \neq 0$ it is useful to portray the potential $V_R(\mathbf{r})$ generating $\mathbf{E}_R$. The form of the $(-e)\{V_R(\mathbf{r}) - V_R(0)\}$ function along several directions is drawn for both lattices in Fig. 2. For clarifying what are the chosen directions and specially the nature of involved electronic orbitals it is convenient to work also with the trigonal basis $\{x_t,y_t,z_t\}$ defined in Fig. 3. Quantities referred to this basis will be denoted by the subscript $t$.

In Fig. 2 the form of $(-e)V_R(\mathbf{r})$ is depicted for directions called $d_0$, $d_1$ and $d_2$. Here $d_1$ and $d_2$ correspond to $<0,0,1>_t$ and $<-1,1,\sqrt{2}> = <\sqrt{3}, \sqrt{2}, 1>_t$ directions, respectively, while $d_0$ refers to a metal-ligand direction corresponding to $<1,0,0>$ type directions in the $\{x,y,z\}$ basis set (Fig. 3). These directions will be useful in the later discussion.

Although, according to $D_3$ symmetry, $\mathbf{E}_R(0) = 0$ for both host lattices the shape of $(-e)V_R(\mathbf{r})$ looks quite different at the ligand region (Fig. 2). For instance, along the $d_1$ direction of $Be_3Si_6Al_2O_{18}$ the $(-e)V_R(\mathbf{r})$ function shows an *increase* of about 1 eV from the origin to the $\mathbf{r} = (0,0,1)_t$ Å point, while for $MgAl_2O_4$ there is a *lessening* of 3 eV. As regards a metal-ligand direction, $d_0$, $(-e)V_R(\mathbf{r})$ decreases but slightly for beryl while it is practically flat for the spinel. The distinct shape of $(-e)V_R(\mathbf{r})$ along $d_1$ is qualitatively consistent with the quite different value of the $\phi_3$ angle for $MgAl_2O_4$ and $Be_3Si_6Al_2O_{18}$. Considering the spinel lattice, if the electron moves from the central position along a $<0,0,1>_t$ direction it is attracted towards the closer plane of three $Al^{3+}$ ions (Fig. 1). By contrast, in beryl the three $Be^{3+}$ ions of the second shell are lying in a plane perpendicular to the $d_1$ direction ($C_3$ axis) and thus $(-e)\{V_R(\mathbf{r}) - V_R(0)\}$ should behave in an opposite way.

The different 10Dq values exhibited by $MgAl_2O_4$:$Cr^{3+}$ and emerald can qualitatively be understood just considering the effects of $(-e)V_R(\mathbf{r})$ upon $e_g$ ($\sim x^2-y^2$, $3z^2-r^2$) and $t_{2g}$ ($\sim xy$, xz, yz) orbitals in first-order perturbation. In fact, at least $\sim 80\%$ of the 10Dq value is already obtained through a calculation of the complex in vacuo (Table 1).

Let us firstly consider the antibonding $e_g$ ($\sim x^2-y^2$, $3z^2-r^2$) orbitals in cubic symmetry. It should be recalled here that although such orbitals transform like $x^2-y^2$ and $3z^2-r^2$ wavefunctions of central cation the actual molecular orbital wavefunctions involves an admixtures with 2p and 2s wavefunctions of oxygen ligands. As the degeneracy in $e_g$ is not removed by a trigonal distortion we can consider that in the present cases such orbitals describe in a first approximation the *e* orbitals in $D_3$ symmetry. Bearing in mind that $e_g$ ($\sim x^2-y^2$, $3z^2-r^2$) orbitals are mainly directed towards ligands ($d_o$ directions) and looking at Fig. 2, it can be expected that $\mathbf{E}_R$ has practically no effect for $MgAl_2O_4$:$Cr^{3+}$ while it would induce a decrease of the energy of such orbitals in the case of the emerald thus favouring a lessening of 10Dq. It should be remarked that, in order to interpret Fig. 2, such effect depends on the probability of finding an $e_g$ electron on ligands and thus on the covalency of the chemical bonding between chromium and oxygen ligands. Present calculations give a total charge on ligands equal to 25% (14%) for an electron in an $e_g$ ($t_{2g}$) orbital.

More interesting effects appear precisely in the case of $t_{2g}$ ($\sim$xy, xz, yz) orbitals in cubic symmetry. First of all the trigonal distortion splits $t_{2g}$ into a singlet $a$ and a doublet $e$. The wavefunction of the $a$ singlet transforms like (xy + xz + yz), that is, $3z_t^2 - r_t^2$ in the trigonal basis (Fig. 3). This means that the $a$ orbital is directed along the $C_3$ axis although some density is also located in the perpendicular plane (Fig. 4). As regards the $e(t_{2g})$ doublet the two orbitals forming the basis can be chosen as (xz – yz) $\sim$ ($\sqrt{2}x_ty_t$ + $x_tz_t$) and (xz + yz - 2xy) $\sim$ ($x_t^2$-$y_t^2$ + $\sqrt{2}y_tz_t$). Considering the (xz – yz) $\sim$ ($\sqrt{2}x_ty_t$ + $x_tz_t$) orbital of the $e(t_{2g})$ doublet it involves an admixture of the $x_ty_t$ orbital, lying in the perpendicular plane to the $C_3$ axis, with the $x_tz_t$ lying outside that plane. The (xz – yz) $\sim$ ($\sqrt{2}x_ty_t$ + $x_tz_t$) orbital possesses four lobes, two placed along $<$-1,1,$\sqrt{2}$> = $<\sqrt{3}$,$\sqrt{2}$,1>$_t$ and two along $<$1,-1,$\sqrt{2}$> = $<$-$\sqrt{3}$, $\sqrt{2}$, 1>$_t$ direction. Bearing in mind these considerations it is possible to understand the different influence of (-$e$)$V_R(\mathbf{r})$ on $t_{2g}$($\sim$xy, xz, yz) orbitals in the two host lattices. Let us first consider the $a(t_{2g})$ orbital. Looking at Figs. 2 and 4 it is clear that in the case of emerald the electronic density lying around $<$0,0,1>$_t$ increases its energy due to the action of $V_R(\mathbf{r})$. By contrast, for the spinel the electronic density located in the neighbourhood of $\mathbf{r}$ = $<$0,0,1.5>$_t$ Å is subject to (-$e$)$\{V_R(\mathbf{r})$- $V_R(0)\}$ ≈ -3 eV, which tends to decrease the energy of the $a(t_{2g})$ orbital. In Fig. 2 is also portrayed the form of $V_R(\mathbf{r})$ along a $d_2$ direction corresponding to one of the lobes of the (xz – yz) $\sim$ ($\sqrt{2}x_ty_t$ + $x_tz_t$) orbital. It can be noticed that for both $MgAl_2O_4$:$Cr^{3+}$ and emerald (-$e$)$V_R(\mathbf{r})$ is practically constant for | $\mathbf{r}$ | < 1.5 Å, although it increases slightly for higher distances. In view of these considerations, it can be expected that the energy of the $t_{2g}$ barycentre of the emerald is increased with respect to that of the spinel due to the action of the internal electric field. This fact helps again to lessen the value of 10Dq in the former case and to enhance it in the latter one. The present argument is thus in qualitative agreement with the calculated values shown in Table 1.

## 3. Final Remarks

The present study shows that the different colour of $MgAl_2O_4$:$Cr^{3+}$ and emerald can be well explained considering the $CrO_6^{9-}$ complex subject to the corresponding internal field $\mathbf{E}_R$. This result is thus consistent with recent findings showing that the shift undergone by crystal–field and charge transitions of $Cr^{3+}$ and $Fe^{3+}$ impurities on passing from beryl to corundum can also be ascribed to the different shape of $E_R$ in the two host lattices[39].

Although the local symmetry around the impurity is the same in $MgAl_2O_4:Cr^{3+}$ and emerald, however the arrangement of close ions and consequently the behaviour of $V_R(\mathbf{r})$ is quite different in both lattices as stressed by Figs. 1 and 2. It is worth remarking here that the importance of $V_R(\mathbf{r})$ in the present problem is enhanced due to the directionality of orbitals. By virtue of this fact, the electronic density in $e_g$ and $t_{2g}$ orbitals is not isotropically distributed in the complex region.

The present calculations gathered in Table 1 lead to a 10Dq value for $MgAl_2O_4:Cr^{3+}$ which is higher than that for ruby. Although this is in qualitative agreement with experiments[1--3,30,31], the observed difference in 10Dq between both systems, $\Delta_{SR}$ is only of 450 cm$^{-1}$, and thus this difference is overestimated by the present calculations leading to $\Delta_{SR} = 2600$ cm$^{-1}$ using $R_I = 1.98$ Å for $MgAl_2O_4:Cr^{3+}$ (Table 1). Apart from the fact that discrepancies between experimental and calculated 10Dq values of about 1000 cm$^{-1}$ are very common, there are two factors that could contribute to reduce this overestimation. On one hand, if there is an experimental uncertainty of ±0.01 Å for each system this could lead to a decrease of 1000 cm$^{-1}$ in $\Delta_{SR}$. On the other hand, the calculated splitting between $e(t_{2g})$ and $a(t_{2g})$ orbitals for $MgAl_2O_4:Cr^{3+}$ is ~2600 cm$^{-1}$ which is not observed experimentally. We have verified that this splitting is greatly due to a 25% contamination of 4s orbitals in the singlet $a(t_{2g})$ which lies below $e(t_{2g})$. If this anomaly is eliminated this would decrease 10Dq by ~600 cm$^{-1}$.

The present results underline that the difference $\Delta_{SE}$ can reasonably be understood considering the effects of the corresponding $V_R(\mathbf{r})$ potential in first-order perturbation. As it has been previously underlined the situation can be more complex in the case of ruby. It has been argued[5] that due to the existence of an electric field along the whole metal-ligand direction it can give rise to more important changes in the electronic density with respect to what is found for the complex in vacuo.

Let us now say a few words on the green colour displayed by the $Cr_2O_3$ pure compound[40] which has the same structure as $Al_2O_3$. Compared to $Al_2O_3:Cr^{3+}$ (10Dq = 18070 cm$^{-1}$), the value 10Dq = 16700 cm$^{-1}$ measured for $Cr_2O_3$ involves a shift $\Delta$(10Dq)

= -1370 cm$^{-1}$. Recent EXAFS measurements[6] give a mean distance $R_I$ = 1.965 ± 0.01 Å for ruby while $R_I$ = 1.98 ± 0.01 Å for $Cr_2O_3$. According to Eq. (2), one would expect that on going from $Al_2O_3$:$Cr^{3+}$ to $Cr_2O_3$ the $(10Dq)_v$ quantity would decrease by ~600 cm$^{-1}$. Nevertheless, this figure is about half the experimental value $\Delta(10Dq)$ = -1370 cm$^{-1}$. It is worth noting however that while in $Al_2O_3$ the charge on aluminium is found[41] to be practically equal to +3 the charge on chromium in $Cr_2O_3$ is expected to be smaller as a result of the covalent bonding which is always present in every transition-metal complex. As shown in Section 2, 14% of the electronic charge associated with an electron in an antibonding $t_{2g}$ orbital is found to be placed on ligands. The expected reduction on the absolute value of metal and oxygen charges on passing from $Al_2O_3$ to $Cr_2O_3$ tends to decrease the value of $|V_R(\mathbf{r}) - V_R(0)|$ and thus the $\Delta_R$ contribution. Along this line, it has been shown[5] that if in the $Al_2O_3$ lattice the cation charge goes from +3 to +2.7 it induces a $\Delta_R$ lessening of 550 cm$^{-1}$. Work on this subject is currently under way.

**Acknowledgments**

The authors would like to thank to A. Juhin who suggested to explore the present problem. The support by the Spanish Ministerio de Ciencia y Tecnología under Project FIS2006-02261 is acknowledged.

| System | $R_H$ | R | 10Dq | | |
|---|---|---|---|---|---|
| | | | in vacuo | under $\mathbf{E}_R$ | Experimental |
| Emerald | 1.906 | 1.975 | 16188 | 15739 | 16130 |
| $MgAl_2O_4$ :$Cr^{3+}$ | 1.930 | 1.980 1.995 | 16336 15828 | 20627 19996 | 18520 |
| Ruby | 1.913 | 1.965 | 16043 | 18179 | 18070 |

**Table 1**. Calculated 10Dq values for the $CrO_6^{9-}$ complex *in vacuo* (at the experimental equilibrium geometry[4,8,9]) and under the internal electric field, $\mathbf{E}_R$, coming from $Be_3Si_6Al_2O_{18}$, $MgAl_2O_4$ and $Al_2O_3$ host lattices. The experimental 10Dq values of these systems[1,4,30,31] are also enclosed. In the case of $MgAl_2O_4$:$Cr^{3+}$, 10Dq is given for the experimental distance[8] ($R_I = 1.98 \pm 0.01$ Å) and also for R = 1.995 Å. In the case of ruby, $R_I$ and $R_H$ mean the average $Cr^{3+}$-$O^{2-}$ and $Al^{3+}$-$O^{2-}$ distance respectively. R and $R_H$ are given in Å while 10Dq in $cm^{-1}$.

**Figure captions**

**Figure 1**. (Color online) $CrO_6^{9-}$ complexes and their surrounding shells of neighbours in (a) the emerald and (b) in the spinel. The meaning of the directions $d_1$ and $d_2$ is explained in the text and in the Fig. 3.

**Figure 2.** (Color online) (a) Electrostatic potential $V_R(\mathbf{r})$ of the rest of the lattice ions on a $CrO_6^{9-}$ complex for the case of emerald (above) and spinel (below), depicted along $d_0$, $d_1$ and $d_2$ directions. The meaning of the three directions is explained in the text and in the Fig. 3.

**Figure 3.** (Color online) Main axis of an octahedral basis {x,y,z} and in a trigonal basis {$x_t, y_t, z_t$} related by $x_t = \frac{1}{\sqrt{2}}(-x+y)$, $y_t = \frac{1}{\sqrt{6}}(-x - y + 2z)$, $z_t = \frac{1}{\sqrt{3}}(x + y + z)$. $d_1$ and $d_3$ directions coincide with $z_t$ and $x_t$, respectively, while $d_2$ is written as $d_2 = -\frac{1}{2}x + \frac{1}{2}y + \frac{1}{\sqrt{2}}z$ in the octahedral basis or equivalently $d_2 = \frac{1}{\sqrt{2}}x_t + \frac{1}{\sqrt{3}}y_t + \frac{1}{\sqrt{6}}z_t$ in the trigonal basis.

**Figure 4.** (Color online) Orbitals belonging to the $t_{2g}$ triplet in a $D_3$ symmetry. (a) Singlet orbital *a*. It is written as (xy + xz + yz) in the octahedral basis and $3z_t^2 - r_t^2$ in the trigonal basis. (b) First of the orbitals of the doublet *e*. It is written as (xz – yz) in the octahedral basis and ($\sqrt{2}x_t y_t + x_t z_t$) in the trigonal basis. (c) Second of the orbitals of the doublet *e*. It is written as (xz + yz - 2xy) in the octahedral basis and ($x_t^2 - y_t^2 + \sqrt{2}\, y_t z_t$) in the trigonal basis. The meaning of the directions $d_1$ and $d_2$ is explained in the text and in the Fig. 3.

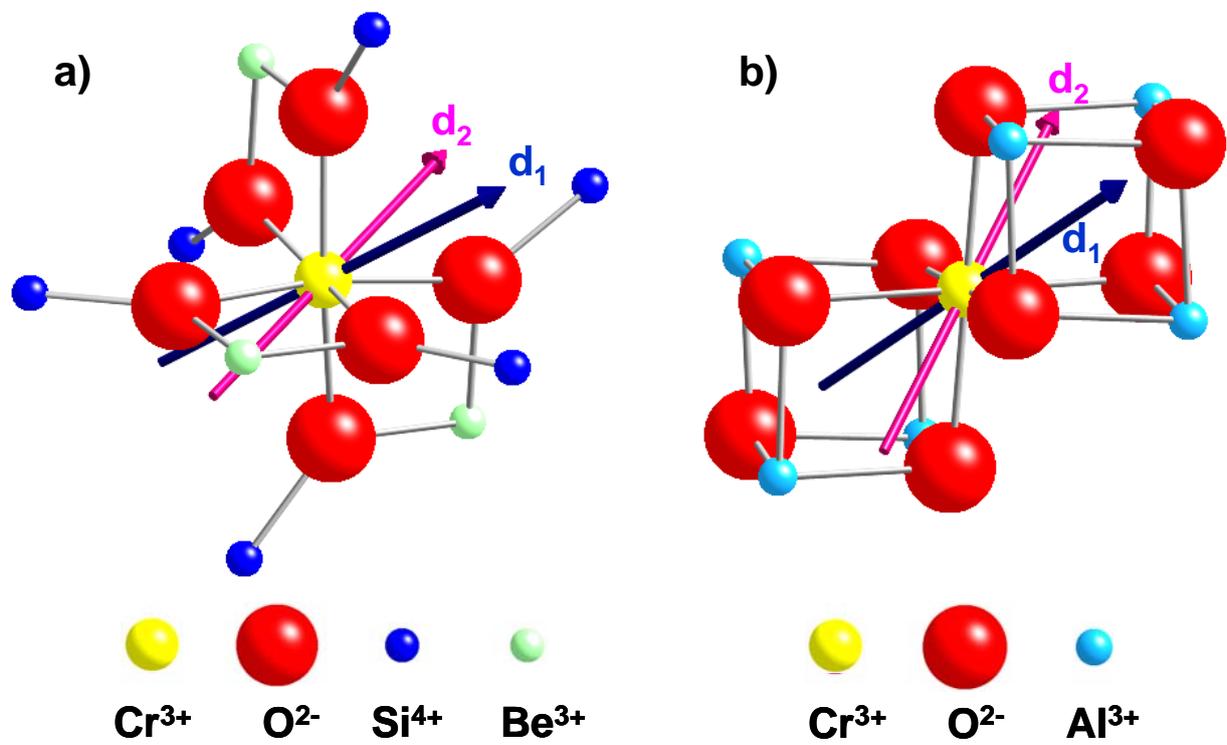

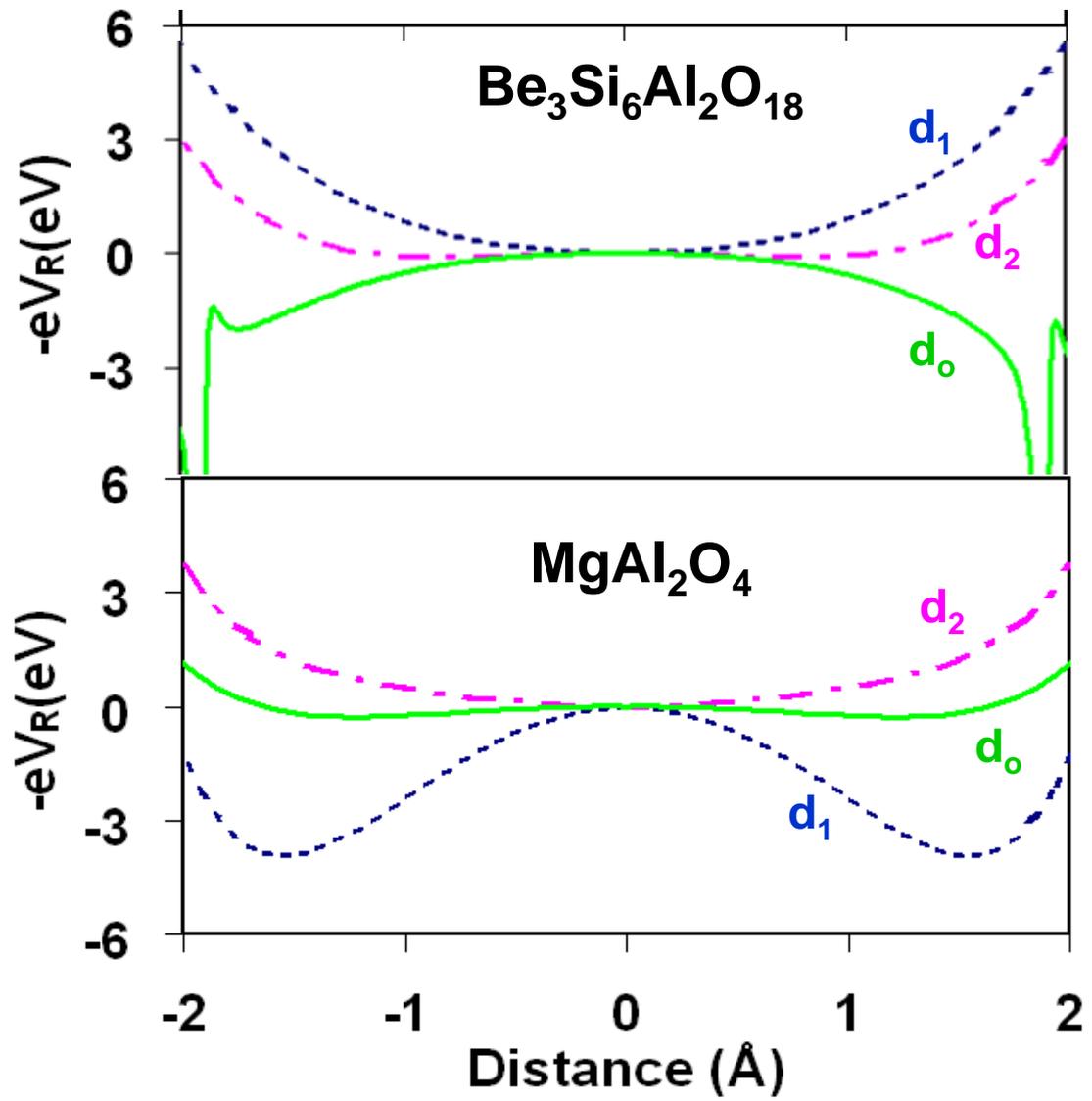

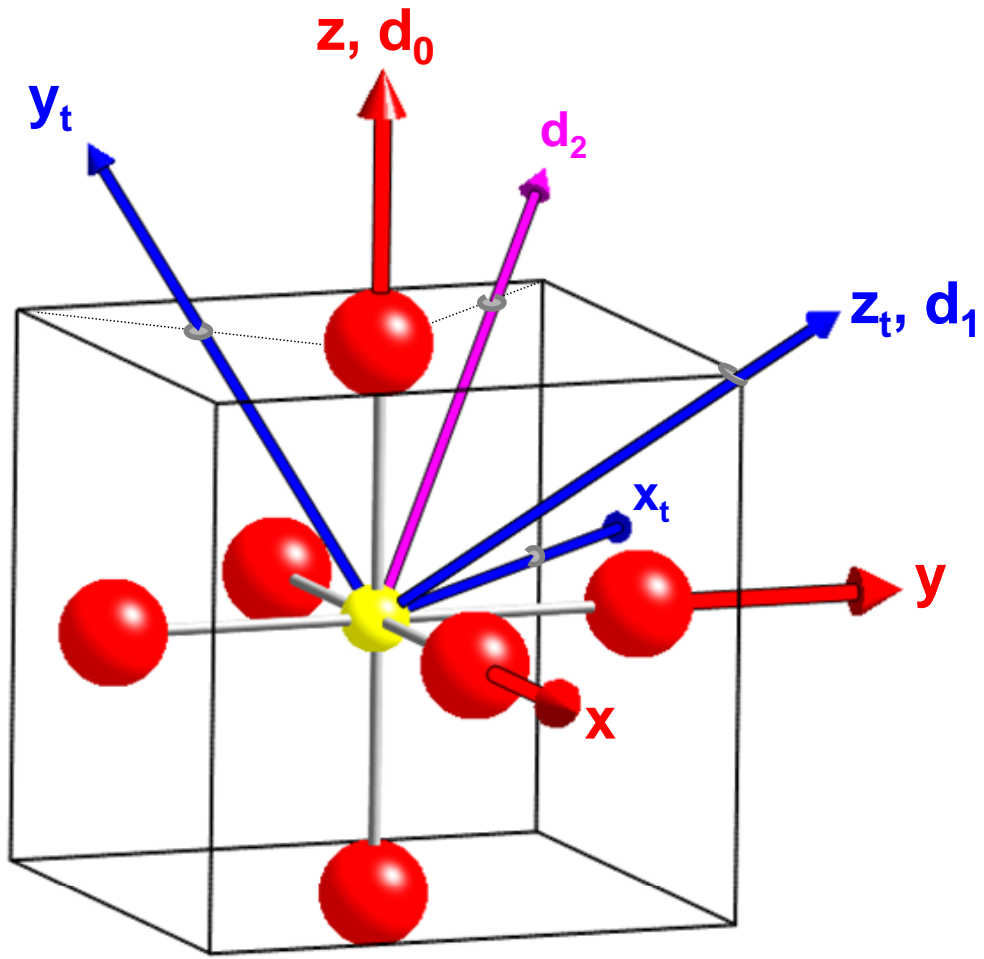

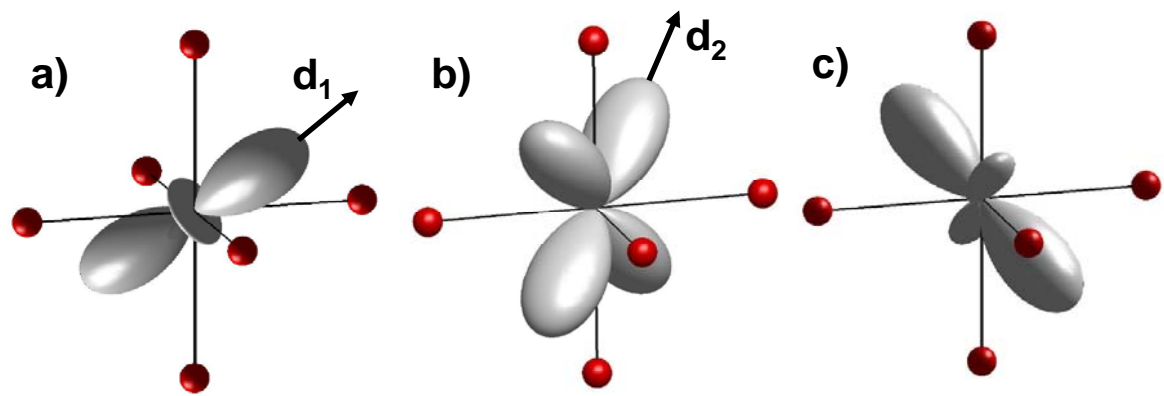